\documentstyle[aps,epsfig,twocolumn]{revtex}

\begin{document}
\title{Experimental Teleportation of a Quantum Controlled-NOT Gate}
\author{Yun-Feng Huang$^{\text{1}}$, Xi-Feng Ren$^{\text{1}}$, Yong-Sheng Zhang$^{%
\text{1}}$, Lu-Ming Duan$^{\text{2,1}}$, and Guang-Can Guo$^{\text{1}}$}
\address{$^{\text{1}}$Laboratory of Quantum Information, University of Science and\\
Technology of China, Hefei, Anhui 230026, P. R. China\\
$^{\text{2}}$Department of Physics and FOCUS center, University of Michigan,%
\\
Ann Arbor, MI 48109}
\maketitle

\begin{abstract}
Teleportation of quantum gates is a critical step for implementation of
quantum networking and teleportation-based models of quantum computation. We
report an experimental demonstration of teleportation of the prototypical
quantum controlled-NOT (CNOT) gate. Assisted with linear optical
manipulations, photon entanglement produced from parametric down conversion,
and coincidence measurements, we teleport the quantum CNOT gate from acting
on local qubits to acting on remote qubits. The quality of the quantum gate
teleportation is characterized through the method of quantum process
tomography, with an average fidelity of $0.84$ demonstrated for the
teleported gate.

\end{abstract}

Physical implementation of quantum computation requires coherent
manipulation of a large number of quantum bits. To scale up the number of
qubits in real physical systems, a particularly interesting approach is to
connect individual physical setups together through some quantum
communication channels and perform distributed quantum computation over all
the nodes \cite{Cirac,Din}. Such a quantum networking approach has provided
practical scaling methods for several promising candidate systems for
implementation of quantum computation \cite{1,2}. For instance, in trapped
ion or cavity quantum-electro-dynamical (QED) systems, the number of qubits
inside an individual trap or cavity could be limited from some practical
considerations, but the limitation can be overcome by wiring up those
individual systems through photon connection \cite{1,2}.

Distributed quantum computation requires one to perform collective quantum
gates on remote qubits. The best way to achieve that is through quantum
teleportation \cite{3}, i.e., one can teleport a collective quantum gate
from acting on local qubits to acting on remote qubits \cite{4,5,2}.
Teleportation of quantum states has been demonstrated in several physical
systems \cite{Tel}, from a photon to a photon, or from an atom to an atom.
However, teleportation of collective quantum gates is more challenging than
teleportation of quantum states. For instance, if one wants to achieve a
remote quantum controlled-NOT (CNOT) gate by teleporting the quantum state
back and forth, one needs two rounds of state teleportations and a local
CNOT gate, which consumes two ebits (entanglement bits), four cbits
(classical bits), and several local collective operations. A better way to
achieve nonlocal quantum CNOT gate on remote qubits is through direct
teleportation of quantum gates \cite{4,2}. The minimum communication cost
for teleportation of a quantum CNOT gate is one ebit and two cbits \cite{5}.

Here, we report an experiment which demonstrates complete teleportation of
the prototypical quantum CNOT gate on photonic qubits. Through linear
optical manipulation and with assistance of entanglement generated from
spontaneous parametric down conversion (SPDC), we teleport a local CNOT
gate, which acts on polarization and path qubits of a single photon, to a
remote CNOT gate, acting on polarization qubits of two distant photons. The
quality of the quantum gate teleportation is characterized through quantum
process tomography \cite{Chuang}, which allows one to fully construct the
teleported quantum gate operation by measurement of its effects on a series
of basis states. Through this method, we demonstrate that the teleported
quantum CNOT gate has a mean fidelity of $0.84$, averaged over all the input
states. With linear optical manipulations, quantum CNOT gates on different
photons have also been directly demonstrated in the coincidence basis by
several recent experiments \cite{Pittman,Gaspa}. The demonstration of
teleportation of the quantum CNOT gate, apart from its fundamental interest,
is a significant step towards realization of quantum networking \cite{1,2}
and teleportation-based models of quantum computation \cite{Gott,Rauss,Knill}%
.

First, we briefly explain the basic idea of teleportation of quantum gate
operations \cite{4,5,2}. Assume that we have two parties, Alice and Bob,
each having two qubits 1,2 and 3,4 (see Fig. 1). We have a shared EPR state $%
\left| \Phi \right\rangle _{23}=\frac 1{\sqrt{2}}(\left| 00\right\rangle
_{23}+\left| 11\right\rangle _{23})$ between the qubits 2 and 3, and our
purpose is to perform a nonlocal quantum CNOT gate on the qubits 1 and 4
with assistance of this shared EPR state, local gate operations on qubits
1,2 and 3,4, and classical communications. The input state of the qubits 1,4
is arbitrary, expressed in general as $\left| \Psi \right\rangle
_{14}=d_0\left| 00\right\rangle _{14}+d_1\left| 01\right\rangle
_{14}+d_2\left| 10\right\rangle _{14}+d_3\left| 11\right\rangle _{14}$. To
achieve the goal, we first perform a quantum CNOT gate $C_{12}$ (where the
first subscript of $C$ denotes the control qubit and the second is the
target qubit) on the local qubits 1,2, and then teleport this gate from the
qubits 1,2 to the qubits 1,4 through the gate teleportation. The gate
teleportation is possible because of the following identity \cite{2}

\begin{eqnarray}
C_{34}C_{12}\left( \left| \Psi \right\rangle _{14}\otimes \left| \Phi
\right\rangle _{23}\right) &=&\left| 0+\right\rangle _{23}\otimes
C_{14}\left( \left| \Psi \right\rangle _{14}\right)  \nonumber \\
&&+\left| 0-\right\rangle _{23}\otimes \sigma _1^zC_{14}\left( \left| \Psi
\right\rangle _{14}\right)  \nonumber \\
&&+\left| 1+\right\rangle _{23}\otimes \sigma _4^xC_{14}\left( \left| \Psi
\right\rangle _{14}\right) \\
&&+\left| 1-\right\rangle _{23}\otimes \left( -\sigma _1^z\sigma _4^x\right)
C_{14}\left( \left| \Psi \right\rangle _{14}\right) .  \nonumber
\end{eqnarray}
where $\left| \pm \right\rangle _3=\left( \left| 0\right\rangle _3\pm \left|
1\right\rangle _3\right) /\sqrt{2}$, and $\sigma _1^z$ and $\sigma _4^x$ are
the Pauli operators acting on the corresponding qubits. This identity shows
that teleportation of the CNOT gate $C_{12}$ is achieved as follows
(illustrated in Fig. 1): first we apply a CNOT gate on the local qubits 3,4,
and then measure the qubits 2,3 respectively in the basis $\left\{ \left|
0\right\rangle _2,\left| 1\right\rangle _2\right\} $ and $\left\{ \left|
+\right\rangle _3,\left| -\right\rangle _3\right\} $. Conditional on the
measurement outcomes, we perform one of the following single-bit corrections 
$\left\{ I,\sigma _1^z,\sigma _4^x,-\sigma _1^z\sigma _4^x\right\} $ on the
qubits 1,4. The whole procedure teleports the quantum CNOT\ gate from the
local qubits 1,2 to the remote qubits 1,4 ($C_{12}\rightarrow C_{14}$). The
teleportation consumes one ebit represented by the state $\left| \Phi
\right\rangle _{23}$, and one cbit in each direction to communicate the
measurement outcomes. This procedure has the minimum communication cost for
a remote quantum CNOT\ gate \cite{5}.

\begin{figure}[tb]
\epsfig{file=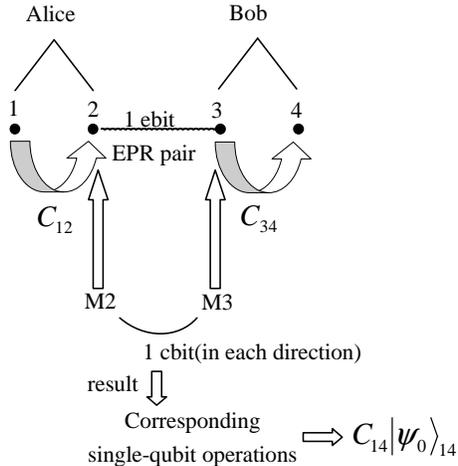,width=6cm} \caption{A sketch to show the idea of teleportation of the quantum CNOT
gate. Each dot stands for one qubit. M2 and M3 represent single qubit
measurements of qubit 2 and 3 in an appropriate basis, and the measurement
outcomes are communicated between Alice and Bob.}
\label{fig1}
\end{figure}

To experimentally demonstrate teleportation of the quantum CNOT\ gate, we
need to have the resource for standard teleportation as well as the ability
of performing CNOT\ gates on local qubits. Similar to the state
teleportation experiment by the De Martini's group \cite{Tel}, we choose the
local qubits represented by the polarization and the path degrees of freedom
of a single photon. This allows us to easily perform deterministic CNOT\
gates on the local qubits through linear optical manipulation. That ability,
together with photon entanglement produced from the SPDC setup, allows us to
teleport the deterministic CNOT from two local qubits carried by a single
photon to nonlocal qubits represented by two remote photons.

Our experimental setup is shown schematically in Fig. 2. A $0.59$ mm thick
BBO ($\beta -B_{a}B_{2}O_{4}$) crystal arranged in a Kwiat-type
configuration \cite{Kwiat} is pumped by a $351.1$ nm laser beam ($100$ mW)
produced by an $Ar^{+}$ laser (Coherent, Sabre, model DBW25/7). Through the
SPDC process, photon pairs are generated in an polarization-entangled EPR
state $(\left| HH\right\rangle +\left| VV\right\rangle )/\sqrt{2}$, where $H$
and $V$ stand for two orthogonal linear polarizations. To introduce the path
qubit, we split each out beam of BBO with a polarizing beam splitter (PBS1
and PBS2), which transmits the horizontally polarized photon ($\left|
H\right\rangle $) and reflects the vertically polarized photon ($\left|
V\right\rangle $). The two paths after the PBS are denoted by $\left|
0\right\rangle $ and $\left| 1\right\rangle $, respectively. We use a half
wave plate (HWP1 and HWP2 in Fig. 2) in the path $\left| 1\right\rangle $ to
exchange the polarization states $\left| V\right\rangle $ and $\left|
H\right\rangle .$ After that, the polarization entanglement is transferred
to the path entanglement, with the whole state having the form

\begin{equation}
\left| \Psi \right\rangle _{1234}=\left| H\right\rangle _{1}\left[ (\left|
00\right\rangle _{23}+\left| 11\right\rangle _{23})/\sqrt{2}\right] \left|
H\right\rangle _{4}\text{,}
\end{equation}
where the subscript denotes different qubits. The qubits 1,2 and 3,4 are
carried respectively by the first and the second photons.

\begin{figure}[tb]
\epsfig{file=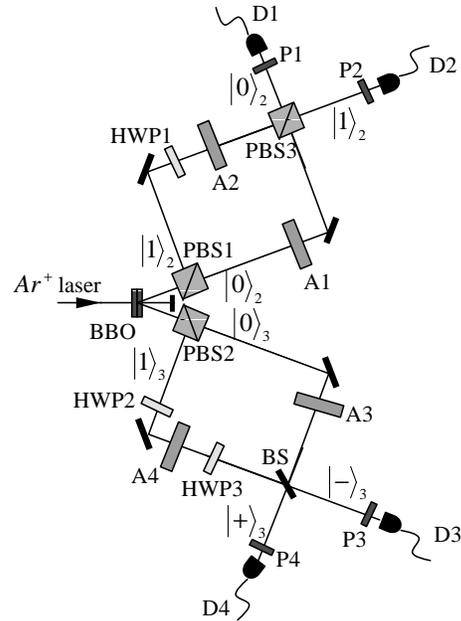,width=6cm} \caption{Schematic experimental setup for teleportation of the quantum CNOT
gate. Qubits 1,2 and 3,4 are carried respectively by the polarization and
the path degrees of freedom of the upper (1,2) and the lower (3,4) photons.
Each of the elements A1-A4 consists of a HWP and a QUWP, used for
preparation of arbitrary polarization states. Each of the polarization
analyzers P1-P4 also consists of a HWP and a QUWP, which, combined with
single photon detectors, measures the photon polarization state in an
arbitrary basis.}
\label{fig2}
\end{figure}

To demonstrate teleportation of the CNOT gate, we need to measure its
effects for an arbitrary input state of the qubits 1 and 4. The qubits 1 and
4 can be prepared in arbitrary polarizations by rotating from the state $%
\left| H\right\rangle $ with a HWP and a quarter wave plate (QUWP), which
are denoted in Fig. 2 as A1, A2, A3, A4 in each path of the photons. The
local CNOT gate $C_{12}$ is achieved by interfering the $\left|
0\right\rangle _2$ and $\left| 1\right\rangle _2$ paths at a PBS (PBS3),
which change the path of photon if it is in the V polarization \cite{Cerf}.
To have phase stability, the M-Z interferometer formed by PBS1 and PBS3 is
equal-armed. The other local CNOT $C_{34}$ is simply realized by putting a
HWP (HWP3, set at $45^{\circ }$) in the $\left| 1\right\rangle _3$ path, to
reverse the polarization state of qubit 4 if qubit 3 is in the state $\left|
1\right\rangle $.

The next step is to perform measurement of qubit 2 \ and 3 respectively in
the basis $\{\left| 0\right\rangle ,\left| 1\right\rangle \}$ and $\{\left|
+\right\rangle ,\left| -\right\rangle \}$. The measurement of qubit 2 is
easily done by detecting two outputs of PBS3 with a single-photon detector.
The measurement of qubit 3 requires another interference of the photon paths 
$\left| 0\right\rangle _{3}$ and $\left| 1\right\rangle _{3}$ at a $50/50$
beam splitter (BS) before the single-photon detection. The M-Z
interferometer formed by PBS2 and BS is also equal-armed. We then register
the four possible coincidences D1-D4, D1-D3, D2-D4, D2-D3 from the four
single-photon detectors (D1 to D4), corresponding detection of the states $%
\left| 0\right\rangle _{2}\left| +\right\rangle _{3}$, $\left|
0\right\rangle _{2}\left| -\right\rangle _{3}$,$\left| 1\right\rangle
_{2}\left| +\right\rangle _{3}$, and $\left| 1\right\rangle _{2}\left|
-\right\rangle _{3}$, respectively. Before each single-photon detector, we
insert a polarization analyzer (P1 to P4). By recording the change of
coincidence counts with rotation of P1-P4, we construct the final
polarization state of qubits 1 and 4 through quantum state tomography \cite
{James}.

\begin{figure}[tb]
\epsfig{file=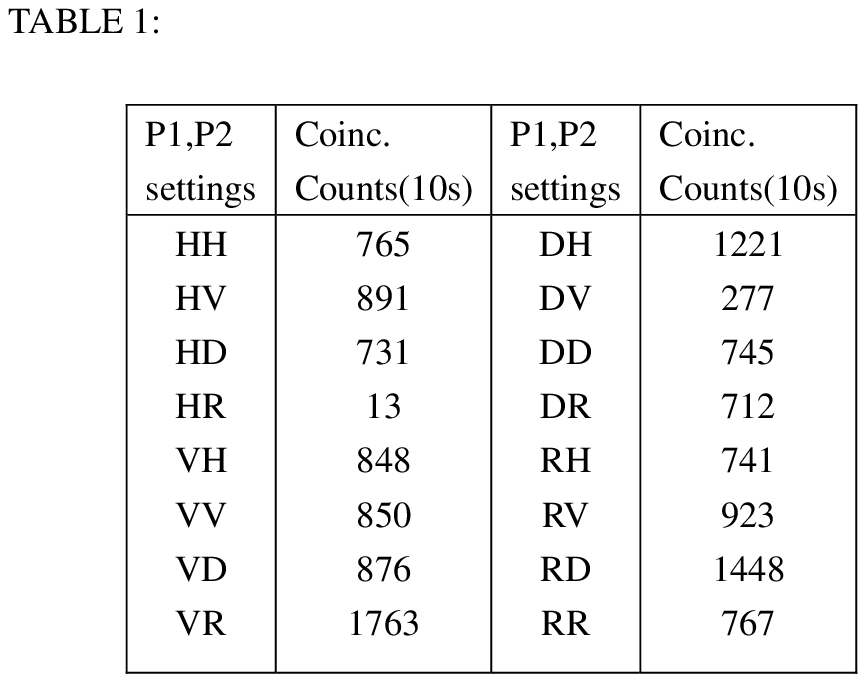,width=6cm} \caption{The coincidence counts of the detectors D1 and D4 for a period of $%
10$ seconds with $16$ different settings of the polarization analyzers P1
and P4 . The input state of qubits 1 and 4 is given by $\left|
R\right\rangle _{1}\left| R\right\rangle _{4}$. The corresponding density
matrix constructed from these data is shown in Fig. 4a.}
\label{fig3}
\end{figure}

To confirm that the local gate $C_{12}$ has been teleported to a remote gate 
$C_{14}$, we verify that the teleported gate generates entanglement between
qubits 1 and 4 from a product input state $\left| \Psi _{in}\right\rangle
_{14}=\left| R\right\rangle _1\left| R\right\rangle _4$, where $\left|
R\left( L\right) \right\rangle =\left( \left| H\right\rangle +\left(
-\right) i\left| V\right\rangle \right) /\sqrt{2}$. If teleportation is
perfect, we should get an entangled output state $\left| \Psi
_{out}\right\rangle _{14}=C_{14}\left| \Psi _{in}\right\rangle _{14}=\left(
\left| H\right\rangle _1\left| R\right\rangle _4-\left| V\right\rangle
_1\left| L\right\rangle _4\right) /\sqrt{2}$. We construct the real density
matrix $\rho _{14}^{%
\mathop{\rm real}%
}$ of qubits 1 and 4 by measuring the coincidences D1-D4 with $16$ different
settings of the polarization analyzers P1 and P4, with the results shown in
Table 1. From the data, with the standard method of quantum state tomography 
\cite{James}, we construct the measured density matrix $\rho _{14}^{%
\mathop{\rm real}%
}$, with its real and imaginary elements shown in Fig. 4. The results are
compared with those from the ideal density matrix $\rho _{14}^{ideal}=\left|
\Psi _{out}\right\rangle _{14}\left\langle \Psi _{out}\right| $. We find
from the data that the state fidelity $F_s\equiv _{14}\left\langle \Psi
_{out}\right| \rho _{14}^{%
\mathop{\rm real}%
}\left| \Psi _{out}\right\rangle _{14}\simeq 0.81$, which significantly
exceeds the criterion of $F_s=0.5$ for demonstration of entanglement of the
state $\rho _{14}^{%
\mathop{\rm real}%
}$ \cite{Bo}. With similar methods, we have also measured the output state
fidelity when qubits 1,4 are input with one of the basis states $\left|
HH\right\rangle $, $\left| HV\right\rangle $, $\left| VH\right\rangle $, $%
\left| VV\right\rangle $, with the results given respectively by $%
F_s^{HH}=0.97$, $F_s^{HV}=0.97$, $F_s^{VH}=0.99$, $F_s^{VV}=0.98$.

\begin{figure}[tb]
\epsfig{file=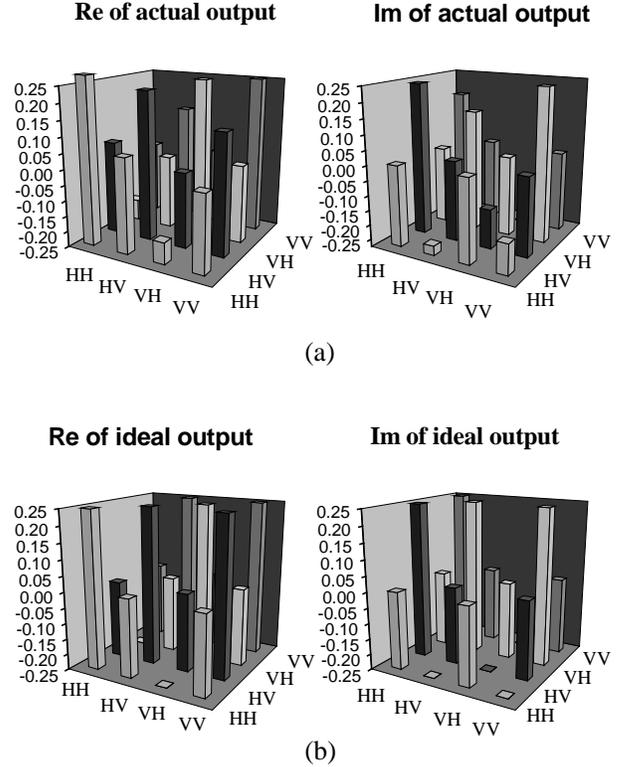,width=8cm}\caption{(a). The real and the imaginary parts of the output density matrix
reconstructed from the coincidence data in Table 1. (b). The corresponding
density matrix elements in the ideal case with a CNOT gate acting on the
input state $\left| R\right\rangle _{1}\left| R\right\rangle _{4}$.}
\label{fig4}
\end{figure}

To fully characterize the teleported quantum gate, one would like to
construct from measurements the corresponding super-operator through the
method of quantum process tomography (QPT) \cite{White,Brien}. The QPT of
the quantum CNOT gate requires us to measure the output density matrices
corresponding to $16$ different input states of qubit 1 and 4 \cite{Brien},
respectively given by $\left| HH\right\rangle $, $\left| HV\right\rangle $, $%
\left| HD\right\rangle $, $\left| HR\right\rangle $, $\left| VH\right\rangle 
$, $\left| VV\right\rangle $, $\left| VD\right\rangle $, $\left|
VR\right\rangle $, $\left| DH\right\rangle $, $\left| DV\right\rangle $, $%
\left| DD\right\rangle $, $\left| DR\right\rangle $, $\left| RH\right\rangle 
$, $\left| RV\right\rangle $, $\left| RD\right\rangle $, $\left|
RR\right\rangle $, where $\left| D\right\rangle =(\left| H\right\rangle
+\left| V\right\rangle )/\sqrt{2}$. The reconstruction of each output
density matrix requires 16 coincidence measurements. So, in total we get $%
16\times 16$ coincidence data. The quality of a quantum gate is measured by
the so-called process fidelity $F_P$ (or called gate fidelity), which is
defined as the overlap of the matrices corresponding respectively to the
measured and the ideal super-operators \cite{Brien}. From the $16\times 16$
coincidence data \cite{Ref}, we find $F_P\simeq 0.80$ for our experiment. An
interesting result deduced from the process fidelity $F_P$ is the average
gate fidelity $\overline{F}$, which is defined as the state fidelity between
the measured and the ideal outputs of the CNOT gate, averaged over all
possible input states. There is a simple relation between $\overline{F}$ and 
$F_P$, given by $\overline{F}=(dF_P+1)/(d+1),$ where the dimension $d=4$ for
the CNOT gate \cite{Brien}. So, we conclude that $\overline{F}\simeq 0.84$
for our experiment.

The measured fidelity in our experiment is pretty high. The remaining
imperfection comes from two main sources. The first contribution is from
imperfection of the two M-Z interferometers, whose visibility is about $85\%$%
. The second source is that the EPR state generated from SPDC is not
perfect. We measured the visibility of the polarization entanglement in the
two-photon state right after SPDC, which is as high as $98.2\%$; but through
the state tomography analyses, we find there are still non-negligible
unwanted elements arising from the measured density matrix compared with the
ideal EPR state, which affect our experiment result.

In summary, we have demonstrated teleportation of the quantum CNOT gate by
using photon entanglement generated from SPDC and linear optical
manipulations. We use dual qubit representation for each single-photon,
which allows us to perform local deterministic CNOT gate. This local gate is
then teleported to a remote gate acting on two distant photons. The quality
of the gate teleportation is characterized through the comprehensive quantum
state and process tomography techniques, and an average teleported gate
fidelity as high as $0.84$ is confirmed from the measured data.

This work was supported by the Chinese National Fundamental Research Program
(2001CB309300), the National Natural Science Foundation (No. Grant
60121503), the NSF of China (10304017), the Innovation funds from Chinese
Academy of Sciences. LMD\ acknowledges supports from the ARDA under ARO
contract, the FOCUS seed funding, and the A. P. Sloan Fellowship.

\end{document}